\documentclass[10pt]{article}
\usepackage{OSAmeet}

\usepackage{amsfonts,amsmath,amssymb,amscd}
\usepackage{epsfig}

\newcommand{\ket}[1]{| #1 \rangle}
\newcommand{\bra}[1]{\langle #1 |}

\renewcommand{\phi}{\varphi}

\begin{document}
\title{\textbf{Quantum Networks for Generating Arbitrary Quantum States}}

\author{Phillip Kaye and Michele Mosca}

\address{Department of Combinatorics \& Optimization, University of Waterloo, Waterloo, ON, N2T 2L1,
Canada.}

\email{prkaye@cacr.math.uwaterloo.ca,
mmosca@cacr.math.uwaterloo.ca}

\HeaderAuthorTitleMtg{Kaye et.al.}{Arbitrary States}{ICQI 2001}

\begin{abstract}
Quantum protocols often require the generation of specific quantum
states. We describe a quantum algorithm for generating any
prescribed quantum state. For an important subclass of states,
including pure symmetric states, this algorithm is efficient.
\end{abstract}

\section{Introduction}
Many results in quantum information theory require the generation
of specific quantum states, such as EPR pairs, or the
implementation of specific quantum measurements, such as a von
Neumann measurement in a Fourier transformed basis. Some states and
measurements can be efficiently implemented using standard quantum
computational primitives such as preparing a qubit in the state
$\ket{0}$ and applying a sequence of quantum gates (from a finite
set). EPR pairs can be prepared from the state $\ket{0}\ket{0}$
using a Hadamard gate and a controlled-NOT gate.  A von Neumann
measurement in the Fourier basis can be efficiently realized by
applying an inverse quantum Fourier transform and performing a von
Neumann measurement in the standard computational basis, i.e.
$\{\ket{0},
\ket{1} \}$. However many states and basis changes cannot be
efficiently realized. This paper focusses on the generation of
quantum states. For example, in \cite{HMPEPC98}, their improved
frequency standard experiment requires the preparation of specific
symmetric states on $n$ qubits, where $n$ is a parameter (number of
ions). The algorithm we describe here will efficiently prepare the
required symmetric state.  This short paper will focus on the
algorithm for generating the state, and will ignore issues related
to errors and decoherence (for which the theory of fault-tolerant
error-correction, or other stabilization methods, will apply). We
do not have space to elaborate on the details of precision, but
simple calculations that require $O(\log(\frac{1}{\epsilon}))$
extra space and $polylog(\frac{1}{\epsilon})$ elementary operations
allow us to generate any state with fidelity at least $1-\epsilon$.

Suppose we want to generate the
state $\ket{\Psi}=\sum_{x \in \lbrace 0,1 \rbrace
^n}e^{i\gamma_x}\alpha_x\ket{x}$. In practice it suffices to generate a state
$\ket{\tilde{\Psi}}$ satisfying $|\bra{\tilde{\Psi}}\Psi \rangle|^2
> 1
-
\epsilon$ for a given small real number $\epsilon > 0$. In this
case, it suffices to approximate each $\gamma_x$ and $\alpha_x$ to
accuracy $poly(\epsilon)$ (i.e. $O(log(\frac{1}{\epsilon}))$ bits
of accuracy). Here we have factored out the phase in each term, and
so the $\alpha_x$ are all non-negative real values. Note that if we
can prepare the state $\ket{\widehat{\Psi}}
=
\sum_{x \in \lbrace 0,1 \rbrace ^n}\alpha_x\ket{x}$, then we can
approximate $\ket{\Psi}$ arbitrarily well by introducing
appropriate phase factors using methods discussed in \cite{CEMM98}.
We will therefore focus on a method for generating states
$\ket{\Psi}$ with non-negative real amplitudes.

\section{The algorithm}

In order to create the $n$-qubit state $\ket{\Psi}$, we will
implement in sequence $n$ controlled rotations, with the $k$th
rotation controlled by the state of the previous $k-1$ qubits for
$k > 0$.

We will first define these controlled rotations, and then in the
next section we will describe how we would implement them.

Will extend the definition of $\alpha_x$ to $x
\in
\lbrace 0,1 \rbrace ^j$ for $1\leq j < n$.
Suppose we had a copy of $\ket{\Psi}$, and we measured the leftmost
$j$ qubits in the computational basis. Let $\alpha_x$ be the
non-negative real number so that $\alpha_x^2$ equals the
probability the measurement result is $x$.  Then
$\left(\alpha_{x_1x_2\ldots x_{k-1}0}/\alpha_{x_1x_2\ldots
x_{k-1}}\right)^2$ gives the conditional probability that the $k$th
qubit is $\ket{0}$, conditioned on the state of the first state of
the $k-1$ qubits being $\ket{x_1 x_2 \ldots x_{k-1}}$. Define a
controlled rotation $c-U_{x_1 x_2
\ldots x_{k-1}0}^{\Psi}$ by:
\[\begin{CD}\ket{x_1}\ket{x_2}\ldots\ket{x_{k-1}}\ket{0}
 @>{c-U_{x_1 x_2\ldots x_{k-1}0}^{\Psi}}>>
\ket{x_1}\ket{x_2}\ldots\ket{x_{k-1}}\left(\frac{\alpha_{x_1x_2\ldots
x_{k-1}0}}{\alpha_{x_1x_2\ldots
x_{k-1}}}\ket{0}+\frac{\alpha_{x_1x_2\ldots
x_{k-1}1}}{\alpha_{x_1x_2\ldots x_{k-1}}}\ket{1}\right) \end{CD}\]

\begin{figure}[h]
\begin{center}
\includegraphics[width=2.5in,height=1.5in]{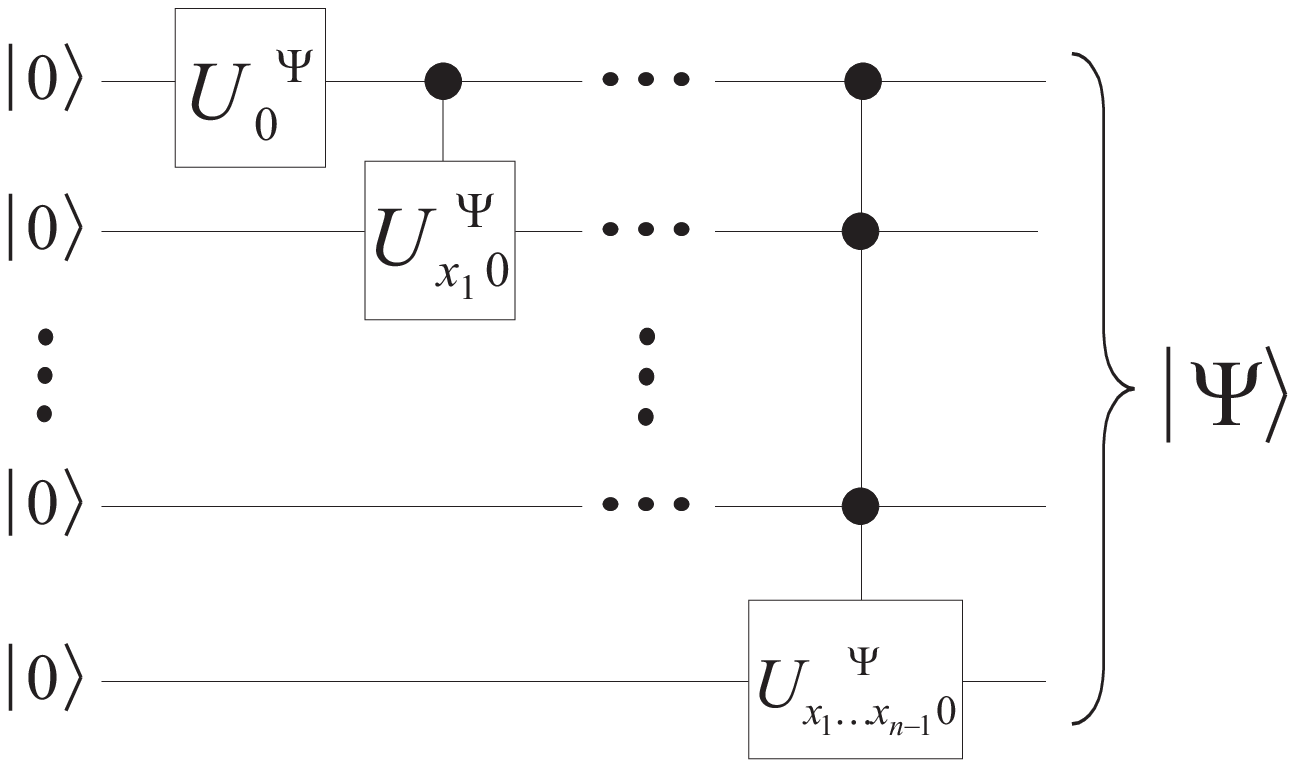}
\end{center}
\caption[]{Network to generate $\ket{\Psi}$} \label{genPsi.fig}
\end{figure}

As shown in Figure \ref{genPsi.fig}, the algorithm for generating
the $n$-qubit state $\ket{\psi}$ is a sequence of $n$ such
controlled rotations. It is easy to show by induction that after
the first $j$ controlled rotations are applied we have produced the
state
\[ \sum_{x_1 x_2 \ldots x_j \in \{0,1\}^j } \alpha_{x_1 x_2 \ldots x_j} \ket{x_1 x_2 \ldots x_j} \]
and therefore after all $n$ controlled rotations we have
\[c-U_{x_1x_2\ldots x_{n-1}0}^\Psi\left(c-U_{x_1x_2\ldots
x_{n-2}0}^\Psi\left(\ldots
c-U_{x_10}^\Psi\left(U_0^\Psi\ket{0}\right)\ket{0}\ldots\right)\ket{0}\right)\ket{0}
=\ket{\Psi} .
\]

\section{Details}

\begin{figure}[h]
\begin{center}
\includegraphics[width=3.5in,height=1.3in]{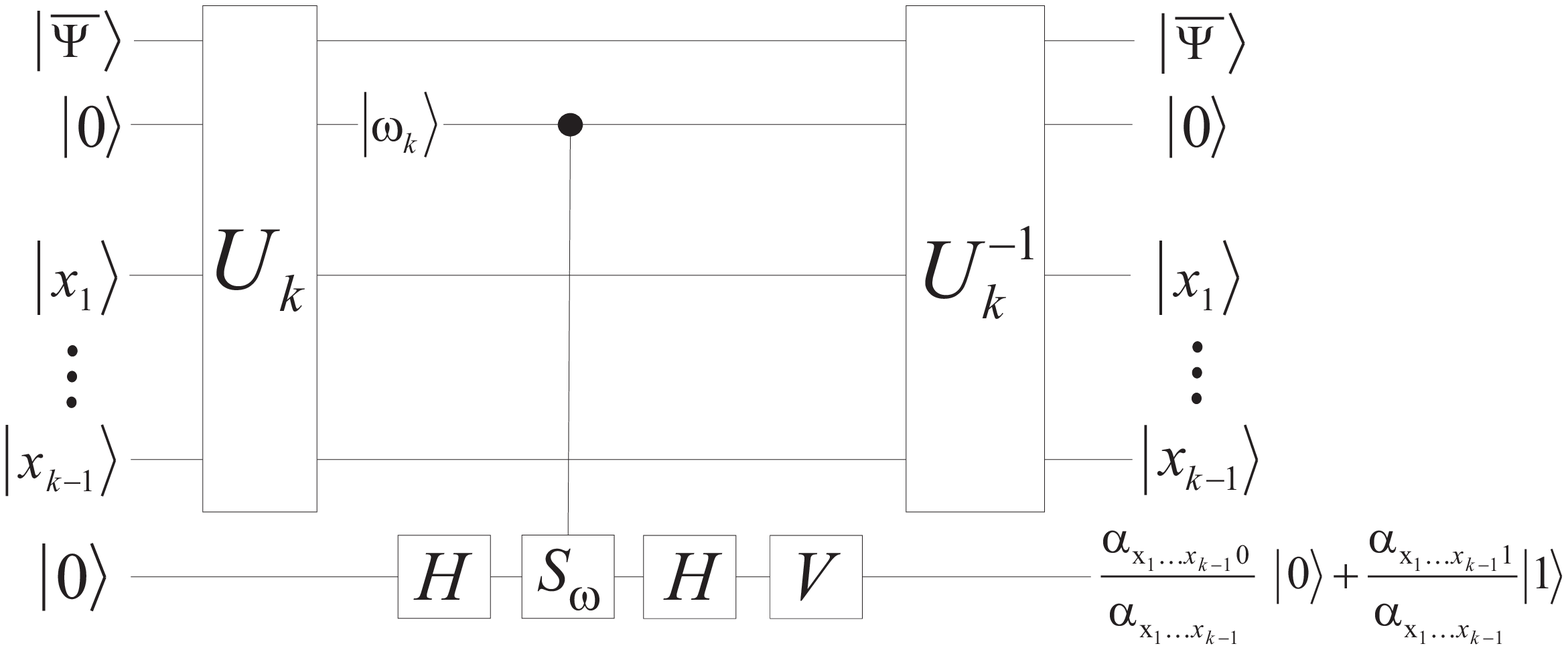}
\end{center}
\caption[]{Network implementing {$c-U_{x_1\ldots
x_{k-1}0}^{\Psi}$}} \label{U_k.fig}
\end{figure}

In this section we show how to implement the
controlled-$U_{x_1\ldots x_{k-1}0}^{\Psi}$ with arbitrary
precision. First assume we have a quantum register
$\ket{\overline{\Psi}}$ which encodes some ``classical''
description of the state $\Psi$. The state $\ket{\overline{\Psi}}$
must contain enough information to allow the probabilities
$\alpha_{x}^2$ (or a related quantity, such as the $\omega_x$ we
define below) to be computed. We also use an ancilla register of
$O(\log(\frac{1}{\epsilon}))$ qubits initialized to the state
$\ket{0}$. Then we define operators $U_k$ for each $1\leq k\leq n$
as follows:
\[
\ket{\overline{\Psi}}\ket{0}\ket{x_1}\ldots\ket{x_{k-1}}
\overset{U_k}{\longrightarrow}
\ket{\overline{\Psi}}\ket{\omega_k}\ket{x_1}\ldots\ket{x_{k-1}}
\]
where $\omega_k$ satisfies
\[\cos^2(2\pi\omega_k)=\left(\alpha_{x_1x_2\ldots
x_{k-1}0}/\alpha_{x_1x_2\ldots x_{k-1}}\right)^2 +
O(poly(\epsilon)).\] A simple application of the techniques in
\cite{CEMM98} allow us to approximate (arbitrarily well) the
transformation:
\[c-S_\omega:\hspace{2mm}
\ket{\omega}\ket{0}\longrightarrow \ket{\omega} e^{2\pi i \omega}\ket{0}
\hspace{1mm},\hspace{1mm} \ket{\omega} \ket{1}\longrightarrow \ket{\omega} e^{-2\pi i
\omega}\ket{1}
.\]
  Also, define
$V=\begin{bmatrix} 1 & 0 \\ 0 & -i\end{bmatrix}$.  With these
components, a network implementing $c-U_{x_1\ldots
x_{k-1}0}^{\Psi}$ is shown in Figure \ref{U_k.fig}.
Here we assumed that the algorithm works for a general family of
states with classical descriptions $\ket{\overline{\Psi}}$. If we
are only interested in producing a specific state $\ket{\Psi}$ the
network can be simplified by removing the register containing
$\ket{\overline{\Psi}}$ and simplifying each $U_k$ to work only for
that specific $\ket{\Psi}$  (in the same way that one can simplify
a circuit for adding variable inputs $x$ and $y$ to one that adds a
fixed input $5$ to variable input $y$).

\section{Efficiency: an example}
The overall efficiency of our algorithm depends on how efficiently
we can implement $U_k$; in other words, how efficiently we can
compute the conditional probabilities $\left(\alpha_{x_1 x_2
\ldots x_k 0}/\alpha_{x_1 x_2 \ldots x_k }\right)^2$ or equivalently
$\left(\alpha_{x_1 x_2
\ldots x_k 1}/\alpha_{x_1 x_2 \ldots x_k }\right)^2$

One example for which this is easy is the symmetric states. The
symmetric state $\ket{S_r}$ is defined to be an equally-weighted
superposition of the computational basis states $\ket{x}$ that
have Hamming weight $H(x)=r$ ($H(x)$ is the number of bits of $x$
that equal 1). That is,
\[\ket{S_r}=\frac{1}{\sqrt{\binom{n}{r}}}\sum_{H(x)=r}\ket{x}.\]
The conditional probability $\left(\frac{\alpha_{x_1 x_2 \ldots
x_{k-1}1}}{\alpha_{x_1 x_2 \ldots x_{k-1} }}\right)^2$ is easily
computed to be
\[\frac{r-H(x_1x_2\ldots x_{k-1})}{n-k}\]
for $1\leq k<n$, and to be $k-H(x_1x_2\ldots x_{k-1})$ for $k=n$.
The Hamming weight can be efficiently computed as shown in
\cite{KM01}. Then we simply need to reversibly compute the
$\omega_k$ satisfying
$\cos^2(2\pi\omega_k)=\left(\frac{\alpha_{x_1x_2\ldots
x_{k-1}0}}{\alpha_{x_1x_2\ldots x_{k-1}}}\right)^2$ +
$poly(\epsilon)$.

Another example for which we can efficiently implement $U_k$ is
for more general symmetric pure states
\[\sum_{j=0}^n \beta_j \ket{S_j}\]
where we are given the $\beta_j$ values (as required
in \cite{HMPEPC98}).

This technique will not allow us to generate efficiently all
quantum states, but it will work for any family of states where
for some reordering of the qubits we can efficiently compute the
conditional probabilities $\left(\frac{\alpha_{x_1 x_2 \ldots x_k
0}}{\alpha_{x_1 x_2 \ldots x_k }}\right)^2$.

\end{document}